\documentstyle[12pt]{article}

\def\be{\begin{equation}}
\def\ee{\end{equation}}

\def\gsim{\mathrel{%
\rlap{\raise 0.511ex \hbox{$>$}}{\lower 0.511ex
\hbox{$\sim$}}}}

\def\lsim{\mathrel{
\rlap{\raise 0.511ex \hbox{$<$}}{\lower 0.511ex
\hbox{$\sim$}}}}

\begin{document}

\vglue 1cm

\centerline{\LARGE \bf Gravity, Equivalence Principle and Clocks}

\vglue 1.5cm

\centerline{{\large Thibault DAMOUR}}

\vglue 1cm

\centerline{Institut des Hautes Etudes Scientifiques, 91440 
Bures-sur-Yvette, France}

\medskip

\centerline{and DARC, CNRS - Observatoire de Paris, 92195
Meudon Cedex, France}

\vglue 2.5cm

\begin{abstract}
String theory suggests the existence of gravitational-strength
scalar fields (``dilaton'' and ``moduli'') whose couplings to
matter violate the equivalence principle. This provides a new
motivation for high-precision clock experiments, as well as a
generic theoretical framework for analyzing their significance.
\end{abstract}

\section{Introduction}

The basic question we wish to address is the following: given
the existing experimental tests of gravity, and given the
currently favored theoretical framework, can high-precision
clock experiments probe interesting theoretical possibilities
which remain yet unconstrained ? In addressing this question
we wish to assume, as theoretical framework, the class of
effective field theories suggested by modern unification
theories, and notably string theory.

To start, let us mention that the theoretical framework most
studied in the phenomenology of gravitation, i.e. the class of
``metric'' theories of gravity \cite{W81}, which includes most
notably the ``Brans-Dicke''-type tensor-scalar theories,
appears as being rather artificial. This is good news because
the phenomenology of ``non metric'' theories is richer and
offers new possibilities for clock experiments. Historically,
the restricted class of ``metric'' theories was introduced in
1956 by Fierz \cite{F56} to prevent, in an {\it ad hoc} way,
too violent a conflict between experimental tests of the
equivalence principle and the existence of a scalar
contribution to gravity as suggested by the theories of
Kaluza-Klein \cite{KK} and Jordan \cite{J}. Indeed, Fierz was
the first one to notice that a Kaluza-Klein scalar would
generically strongly violate the equivalence principle. He
then proposed to restrict artificially the couplings of the
scalar field to matter so as to satisfy the equivalence
principle. The restricted class of
equivalence-principle-preserving couplings introduced by Fierz
is now called ``metric'' couplings. Under the aegis of Dicke,
Nordtvedt, Thorne and Will a lot of attention has been given
to ``metric'' theories of gravity\footnote{Note that
Nordtvedt, Will, Haugan and others (for references see
\cite{W81} and the contributions of Nordtvedt and Haugan to
these proceedings) studied conceivable phenomenological
consequences of generic ``non metric'' couplings, without,
however, using, as we do here, a motivated field-theory
framework describing such couplings.}, and notably to their
quasi-stationary-weak-field phenomenology (``PPN framework'',
see, e.g., \cite{W81}).

By contrast, we wish to stress that nearly all unification
theories (from Kaluza-Klein to string theory) suggest that
``gravitational interactions'' are mediated not only by the
tensor field $(g_{\mu \nu})$ postulated by Einstein, but also
by one or more extra fields, having couplings which violate
the equivalence principle. Among these extra fields the most
universally present seem to be scalar fields, of the type of
the ``dilaton'' of string theory.

Recently, a mechanism has been proposed \cite{DP} to reconcile
in a natural manner the existence of a dilaton field as a
fundamental partner of the graviton field $g_{\mu \nu}$ with
the current level of precision $(\sim 10^{-12})$ of
experimental tests of the equivalence principle. In the
mechanism of \cite{DP} (see also \cite{DN} for
metrically-coupled scalars) the very small couplings necessary
to ensure a near universality of free fall, $\Delta a/a <
10^{-12}$, are dynamically generated by the expansion of the
universe, and are compatible with couplings ``of order unity''
at a fundamental level.

The point of the present paper is to emphasize the rich
phenomenological consequences of dilaton-like fields, and the
fact that high-precision clock experiments might contribute to
searching for, or constraining, their existence.

\section{Generic effective theory of a long-range dilaton}

Motivated by string theory, we consider the generic class of
theories containing a long-range dilaton-like scalar field
$\varphi$. The effective Lagrangian describing these theories
has the form:
\begin{eqnarray}
L_{\rm eff} &=& \frac{1}{4q} R(g_{\mu\nu}) - \frac{1}{2q} \
(\nabla \varphi)^2 - \frac{1}{4e^2 (\varphi)} \ (\nabla_{\mu}
A_{\nu} - \nabla_{\nu} A_{\mu})^2 \nonumber \\
&-& \sum_A \ \left[\overline{\psi}_A \,
\gamma^{\mu} (\nabla_{\mu} -iA_{\mu}) \psi_A + m_A (\varphi)
\, \overline{\psi}_A \psi_A \right] + \cdots \label{eq:01}
\end{eqnarray}
Here, $q\equiv 4\pi \, \overline G$ where $\overline G$ denotes
a bare Newton's constant, $A_{\mu}$ is the electromagnetic
field, and $\psi_A$ a Dirac field describing some fermionic
matter. At the low-energy, effective level (after the breaking
of $SU(2)$ and the confinement of colour), the coupling of the
dilaton $\varphi$ to matter is described by the
$\varphi$-dependence of the fine-structure ``constant'' $e^2
(\varphi)$ and of the various masses $m_A (\varphi)$. Here,
$A$ is a label to distinguish various particles. [A deeper
description would include more coupling functions, e.g.
describing the $\varphi$-dependences of the $U(1)_Y$,
$SU(2)_L$ and $SU(3)_c$ gauge coupling ``constants''.]

The strength of the coupling of the dilaton $\varphi$ to the
mass $m_A (\varphi)$ is given by the quantity
\be
\alpha_A \equiv \frac{\partial \ {\rm ln} \ m_A
(\varphi_0)}{\partial \ \varphi_0} \, , \label{eq:02}
\ee
where $\varphi_0$ denotes the ambient value of $\varphi (x)$
(vacuum expectation value of $\varphi (x)$ around the mass
$m_A$, as generated by external masses and cosmological
history). For instance, the usual PPN parameter $\gamma -1$
measuring the existence of a (scalar) deviation from the pure
tensor interaction of general relativity is given by
\cite{DEF}, \cite{DP}
\be
\gamma -1 = -2 \ \frac{\alpha_{\rm had}^2}{1+\alpha_{\rm had}^2}
\, , \label{eq:03}
\ee
where $\alpha_{\rm had}$ is the (approximately universal)
coupling (\ref{eq:02}) when $A$ denotes any (mainly) hadronic
object.

The Lagrangian (\ref{eq:01}) also predicts (as discussed in
\cite{DP}) a link between the coupling strength (\ref{eq:02})
and the violation of the universality of free fall:
\be
\frac{a_A -a_B}{\frac{1}{2} (a_A + a_B)} \simeq (\alpha_A
-\alpha_B) \alpha_E \sim -5\times 10^{-5} \, \alpha_{\rm
had}^2 \, . \label{eq:04}
\ee
Here, $A$ and $B$ denote two masses falling toward an external
mass $E$ (e.g. the Earth), and the numerical factor $-5 \times
10^{-5}$ corresponds to $A= {\rm Be}$ and $B= {\rm Cu}$. The
experimental limit \cite{Su94}
\be
\left( \frac{\Delta a}{a} \right)_{\rm Be \, Cu} = (-1.9 \pm
2.5) \times 10^{-12} \label{eq:05}
\ee
shows that the (mean hadronic) dilaton coupling strength is
already known to be very small:
\be
\alpha_{\rm had}^2 \lsim 10^{-7} \, . \label{eq:06}
\ee

Free fall experiments, such as Eq. (\ref{eq:05}) or the
comparable Lunar Laser Ranging constraint \cite{LLR}, give the
tightest constraints on any long-range dilaton-like coupling.
Let us mention, for comparison, that solar-system measurements
of the PPN parameters (as well as binary pulsar measurements)
constrain the dilaton-hadron coupling to $\alpha_{\rm had}^2 <
10^{-3}$, while the best current constraint on the time
variation of the fine-structure ``constant'' (deduced from the
Oklo phenomenon), namely \cite{DD96}
\be
-6.7 \times 10^{-17} \, {\rm yr}^{-1} < \frac{d}{dt} \ {\rm
ln} \ e^2 < 5.0 \times 10^{-17} \, {\rm yr}^{-1} \, ,
\label{eq:Oklo}
\ee
yields from Eq. (\ref{eq:16}) below, $\alpha_{\rm had}^2 \lsim
3 \times 10^{-4}$.

To discuss the probing power of clock experiments, we need
also to introduce other coupling strengths, such as
\be
\alpha_{\rm EM} \equiv \frac{\partial \ {\rm ln} \ e^2
(\varphi_0)}{\partial \ \varphi_0} \, , \label{eq:07}
\ee
measuring the $\varphi$-variation of the electromagnetic (EM)
coupling constant\footnote{Note that we do not use the
traditional notation $\alpha$ for the fine-structure constant
$e^2 / 4\pi \hbar c$. We reserve the letter $\alpha$ for
denoting various dilaton-matter coupling strengths. Actually,
the latter coupling strengths are analogue to $e$ (rather than
to $e^2$), as witnessed by the fact that observable deviations
from Einsteinian predictions are proportional to products of
$\alpha$'s, such as $\alpha_A \alpha_E$, $\alpha_{\rm had}^2$,
etc$\ldots$}, and
\be
\alpha_A^{A^*} \equiv \frac{\partial \ {\rm ln} \ E_A^{A^*}
(\varphi_0)}{\partial \ \varphi_0} \, , \label{eq:08}
\ee
where $E_A^{A^*}$ is the energy difference between two atomic
energy levels.

In principle, the quantity $\alpha_A^{A^*}$ can be expressed
in terms of more fundamental quantities such as the ones
defined in Eqs. (\ref{eq:02}) and (\ref{eq:07}). For instance,
in an hyperfine transition
\be
E_A^{A^*} \propto (m_e \, e^4) \ g_I \ \frac{m_e}{m_p} \ e^4 \
F_{\rm rel} (Z e^2) \, , \label{eq:09}
\ee
so that
\be
\alpha_A^{A^*} \simeq 2 \, \alpha_e -\alpha_p + \alpha_{\rm EM}
\left( 4+\frac{d \ {\rm ln} \ F_{\rm rel}}{d \ {\rm ln} \ e^2}
\right) \, . \label{eq:10}
\ee
Here, the term $F_{\rm rel} (Z e^2)$ denotes the relativistic
(Casimir) correction factor \cite{Casimir}. Moreover, in any
theory incorporating gauge unification one expects to have the
approximate link \cite{DP}
\be
\alpha_A \simeq \left( 40.75 - {\rm ln} \ \frac{m_A}{1 \ {\rm
GeV}} \right) \ \alpha_{\rm EM} \, , \label{eq:11}
\ee
at least if $m_A$ is mainly hadronic.

\section{Clock experiments and dilaton couplings}

The coupling parameters introduced above allow one to describe
the deviations from general relativistic predictions in most
clock experiments \cite{TD}. Let us only mention some simple
cases.

First, it is useful to distinguish between ``global'' clock
experiments where one compares spatially distant clocks, and
``local'' clock experiments where the clocks being compared
are next to each other. The simplest global clock experiment
is a static redshift experiment comparing (after transfer by
electromagnetic links) the frequencies of the same transition
$A^* \rightarrow A$ generated in two different locations ${\bf
r}_1$ and ${\bf r}_2$. The theory of Section 2 predicts a
redshift of the form (we use units in which $c=1$)
\be
\frac{\nu_A^{A^*} ({\bf r}_1)}{\nu_A^{A^*} ({\bf r}_2)} \simeq
1 + (1 + \alpha_A^{A^*} \, \alpha_E) \ (\overline{U}_E ({\bf
r}_2) - \overline{U}_E ({\bf r}_1)) \, , \label{eq:12}
\ee
where
\be
\overline{U}_E \, ({\bf r}) = \frac{\overline G \, m_E}{r}
\label{eq:13}
\ee
is the {\it bare} Newtonian potential generated by the
external mass $m_E$ (say, the Earth). Such a result has the
theoretical disadvantage of depending on other experiments for
its interpretation. Indeed, the {\it bare} potential
$\overline{U}_E$ is not directly measurable. The measurement
of the Earth potential by the motion of a certain mass $m_B$
gives access to $(1+\alpha_B \, \alpha_E) \ \overline{U}_E \,
({\bf r})$. The theoretical significance of a global clock
experiment such as (\ref{eq:12}) is therefore fairly indirect,
and involves other experiments and other dilaton couplings. One
can generalize (\ref{eq:12}) to a more general, non static
experiment in which different clocks in relative motion are
compared. Many different ``gravitational potentials'' will
enter the result, making the theoretical significance even
more involved.

A conceptually simpler (and, probably, technologically less
demanding) type of experiment is a differential, ``local''
clock experiment. Such ``null'' clock experiments have been
proposed by Will \cite{W81} and first performed by Turneaure
et al. \cite{T83}. The theoretical significance of such
experiments within the context of dilaton theories is much
simpler than that of global experiments. For instance if
(following the suggestion of \cite{PTM}) one locally compares
two clocks based on hyperfine transitions in alkali atoms with
different atomic number $Z$, one expects to find a ratio of
frequencies 
\be
\frac{\nu_A^{A^*} ({\bf r})}{\nu_B^{B^*} ({\bf r})} \simeq
\frac{F_{\rm rel} (Z_A \, e^2 (\varphi_{\rm loc}))}{F_{\rm
rel} (Z_B \, e^2 (\varphi_{\rm loc}))} \, , \label{eq:14}
\ee
where the local, ambient value of the dilaton field
$\varphi_{\rm loc}$ might vary because of the (relative)
motion of external masses with respect to the clocks
(including the effect of the cosmological expansion). The
directly observable fractional variation of the ratio
(\ref{eq:14}) will consist of two factors:
\be
\delta \ {\rm ln} \ \frac{\nu_A^{A^*}}{\nu_B^{B^*}} = \left[
\frac{\partial \ {\rm ln} \ F_{\rm rel} (Z_A \, e^2)}{\partial
\ {\rm ln} \ e^2} - \frac{\partial \ {\rm ln} \ F_{\rm rel}
(Z_B \, e^2)}{\partial \ {\rm ln} \ e^2} \right] \times \delta
\ {\rm ln} \ e^2 \, . \label{eq:15}
\ee
The ``sensitivity'' factor in brackets due to the
$Z$-dependence of the Casimir term can be made of order unity
\cite{PTM}, while the fractional variation of the
fine-structure constant is expected in dilaton theories to be
of order \cite{DP}, \cite{TD}
\begin{eqnarray}
\delta \ {\rm ln} \ e^2 (t) &=& -2.5 \times 10^{-2} \
\alpha_{\rm had}^2 \ U(t) \nonumber \\
&-& 4.7 \times 10^{-3} \ \kappa^{-1/2} ({\rm tan} \ \theta_0)
\ \alpha_{\rm had}^2 \ H_0 (t-t_0) \, . \label{eq:16}
\end{eqnarray}
Here, $U(t)$ is the value of the externally generated
gravitational potential at the location of the clocks, and
$H_0 \simeq 0.5 \times 10^{-10} \ {\rm yr}^{-1}$ is the Hubble
rate of expansion. [The factor $\kappa^{-1/2} \ {\rm tan} \
\theta_0$ is expected to be $\sim 1$.]

The (rough) theoretical prediction (\ref{eq:16}) allows one to
compare quantitatively the probing power of clock experiments
to that of equivalence principle tests. Let us
(optimistically) assume that clock stabilities of order
$\delta \nu / \nu \sim 10^{-17}$ (for the relevant time scale)
can be achieved. A differential {\it ground} experiment (using
the variation of the Sun's potential due to the Earth
eccentricity) would probe the level $\alpha_{\rm had}^2 \sim
3\times 10^{-6}$. A geocentric satellite differential
experiment could probe $\alpha_{\rm had}^2 \sim 5\times
10^{-7}$. These levels are impressive (compared to present
solar-system tests of the PPN parameter $\gamma$ giving the
constraint $\alpha_{\rm had}^2 \simeq (1-\gamma) / 2 <
10^{-3}$), but are not as good as the present
equivalence-principle limit (\ref{eq:06}). To beat the level
(\ref{eq:06}) one needs to envisage an heliocentric
differential clock experiment (a few solar radii probe within
which two hyper-stable clocks are compared). Such a futuristic
experiment could, according to Eq. (\ref{eq:16}), reach the
level $\alpha_{\rm had}^2 \sim 10^{-9}$. [Let us also note
that a gravitational time delay global experiment using clocks
beyond the Sun as proposed by C. Veillet (SORT concept) might
(optimistically) probe the level $\alpha_{\rm had}^2 \sim
10^{-7}$.] It is, however, to be noted that a much refined test
of the equivalence principle such as STEP (Satellite Test of
the Equivalence Principle) aims at measuring $\Delta a/a \sim
10^{-18}$ which corresponds to the level $\alpha_{\rm had}^2
\sim 10^{-14}$, i.e. five orders of magnitude better than any
conceivable clock experiment.

\section{Conclusions}

In summary, the main points of the present contribution are:
\begin{enumerate}
\item[$\bullet$] Independently of any theory, the result
(\ref{eq:Oklo}) of a recent reanalysis of the Oklo phenomenon
\cite{DD96} gives a motivation, and a target, for improving
laboratory clock tests of the time variation of the
fine-structure constant $e^2$ (which are at the $3.7 \times
10^{-14} \ {\rm yr}^{-1}$ level \cite{PTM}).
\item[$\bullet$] Modern unification theories, and especially
string theory, suggest the existence of new
gravitational-strength fields, notably scalar ones
(``dilaton'' or ``moduli''), whose couplings to matter violate
the equivalence principle. These fields would induce a
spacetime variability of the coupling constants of physics
(such as the fine-structure constant). High-precision clock
experiments are excellent probes of such a possibility.
\item[$\bullet$] The generic class of dilaton theories defined
in Section 2 provides a well-defined theoretical framework in
which one can discuss the phenomenological consequences of the
existence of a dilaton-like field. Such a theoretical
framework (together with some assumptions, e.g. about gauge
unification and the origin of mass hierarchy) allows one to
compare and contrast the probing power of clock experiments to
that of other experiments.
\item[$\bullet$] Local, differential clock experiments (of the
``null'' type of \cite{T83}) appear as conceptually cleaner,
and technologically less demanding, probes of
dilaton-motivated violations of the equivalence principle than
global, absolute clock experiments (of the Gravity Probe A
type).
\item[$\bullet$] If we use the theoretical assumptions of
Section 2 to compare clock experiments to free-fall
experiments, one finds that one needs to send and intercompare
two ultra-high-stability clocks in near-solar orbit in order
to probe dilaton-like theories more deeply than {\it present}
free-fall experiments. Currently proposed improved satellite
tests of the equivalence principle would, however, beat any
clock experiment in probing even more deeply such theories.
\item[$\bullet$] At the qualitative level, it is, however,
important to note that clock experiments (especially of the
``global'', GPA type) probe different combinations of basic
coupling parameters than free-fall experiments. This is
visible in Eq. (\ref{eq:10}) which shows that $\alpha_A^{A^*}$
contains the leptonic quantity $\alpha_e = \partial \ {\rm ln}
\ m_{\rm electron} / \partial \ \varphi_0$ without any small
factor\footnote{Free-fall experiments couple predominantly to
hadronic quantities such as $\alpha_p =
\partial \ {\rm ln} \ m_{\rm proton} / \partial \ \varphi_0$,
and to Coulomb-energy effects proportional to $\alpha_{\rm
EM}$. The effect of the leptonic quantity $\alpha_e$ is down
by a small factor $\sim m_e / m_p \sim 1/1836$.}.
\end{enumerate}

\end{document}